\documentclass{ws-procs961x669}
\usepackage{mathtools}

\newcommand{\D}[0]{\mbox{d}}

\newcommand{\GeneralDerivative}[4]
{ \frac{ {#4}^{#3} {#1} }{ {#4} {#2}^{#3} } }

\newcommand{\pderiv}[2]{ \GeneralDerivative{}{#1}{#2}{\partial} }

\newcommand{\Deriv}[3]{ \GeneralDerivative{#1}{#2}{#3}{\D} }

\newcommand{\ave}[1]{ \left \langle {#1} \right \rangle}
\newcommand{\abs}[1]{ \left| {#1} \right|}

\newcommand{\IM}[1]{\mbox{Im} \left\{ {#1} \right\} }

\newcommand{\ket}[1]{ \left| {#1} \right \rangle }
\newcommand{\bra}[1]{ \left \langle {#1} \right| }
\newcommand{\expect}[3]{\left\langle {#1}\middle|{#2}\middle|{#3}\right\rangle }
\newcommand{\inner}[2]{ \left \langle {#1} \middle| {#2} \right \rangle }
\newcommand{\finner}[2]{ \left\langle {#1}, {#2} \right\rangle }

\newcommand{\subsc}[1]{\mathrm{#1}}

\makeatletter
\renewcommand{\ps@plain}{%
  \renewcommand{\@oddhead}{\hfil{\footnotesize%
    A contribution to the Julian Schwinger Centennial Conference, %
    7--12 February 2018, Singapore}\hfil}%
  \renewcommand{\@evenhead}{\@oddhead}%
  \renewcommand{\@oddfoot}{\hfil\thepage}%
  \renewcommand{\@evenfoot}{\thepage\hfil}%
}
\makeatother
\crop[off]

\begin{document}

\title{Unruh Acceleration Radiation Revisited}

\author{%
  J.~S.~Ben-Benjamin,\textsuperscript{1}
  M.~O.~Scully,\textsuperscript{1,2}
  S.~A.~Fulling, D.~M.~Lee, D.~N.~Page,\textsuperscript{3}
  A.~A.~Svidzinsky,\textsuperscript{1}
  M.~S.~Zubairy}

\address{\centerline{Institute for Quantum Science and Engineering,}
\centerline{Texas A\&M University, College Station, TX 77843, USA}}

\author{%
  M.~J.~Duff,\textsuperscript{4,5}
  R.~Glauber,\textsuperscript{6}
  W.~P.~Schleich,\textsuperscript{2,7}
  W.~G.~Unruh\textsuperscript{8}}

\address{\centerline{Hagler Institute for Advanced Studies,}
\centerline{Texas A\&M University, College Station, TX 77843, USA}}

\address{\begin{minipage}{\textwidth}\centering%
\begin{tabular}{@{}l@{\ }p{0.92\textwidth}@{}}
\textsuperscript{1}&Baylor University, Waco, TX 76706, USA\\
\textsuperscript{2}&Princeton University, Princeton, NJ 08544, USA\\
\textsuperscript{3}&University of Alberta, Edmonton, T6G 2R3, Canada\\
\textsuperscript{4}&Theoretical Physics, Blackett Laboratory, %
Imperial College London, London SW7 2AZ, United Kingdom\\
\textsuperscript{5}&Mathematical Institute, Andrew Wiles Building, %
University of Oxford, Oxford OX2 6GG, United Kingdom\\
\textsuperscript{6}&Harvard University, Cambridge, MA 02138, USA\\
\textsuperscript{7}&Universit\"at Ulm, D-89069 Ulm, Germany\\
\textsuperscript{8}&University of British Columbia, Vancouver, V6T 2A6, Canada
\end{tabular}\end{minipage}}

\begin{abstract}
When ground-state atoms are accelerated and the field with which they interact
is in its normal vacuum state, the atoms detect Unruh radiation.
We show that atoms falling into a black hole emit acceleration radiation
which, under appropriate initial conditions (Boulware vacuum), has an energy
spectrum which looks much like Hawking radiation.
This analysis also provides insight into the Einstein principle of equivalence
between acceleration and gravity.
The Unruh temperature can also be obtained by using the
Kubo--Martin--Schwinger (KMS) periodicity of the two-point thermal correlation
function, for a system undergoing uniform acceleration;
as with much of the material in this paper, this known result is obtained with
a twist.
\end{abstract}

\section*{Ia. Introduction: Dedication}
Julian Schwinger, that towering figure of 20th century physics,
taught us how to tame the infinities of quantum field theory and much more.
For example, he and his students taught us how to profitably apply the
formalism of quantum field theory to the problem of nonequilibrium quantum
statistical mechanics;\cite{kuboMS,schwinger1} yielding, among other things,
the famous KMS condition, which we use herein. 
Indeed, modern quantum optics owes much to Schwinger's Green's
function-correlation function approach.   
In particular, we have found that the tools of quantum optics provide another
window into the problem of Unruh--Hawking radiation. 
It is therefore fitting that we summarize and extend our work on acceleration
radiation in this Schwinger centennial collection.

\begin{figure}[t]
\centerline{\includegraphics[scale=1]{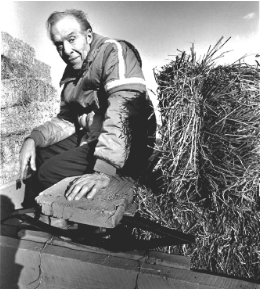}}
\caption{Julian Schwinger rides a hay wagon at the New Mexico Scully ranch in
  1987.}
\end{figure}

\section*{Ib. Introduction: Overview}
The existence of black holes (BHs), regions of spacetime that nothing --- not
even light --- can escape from, is one of predictions of Einstein's general
relativity.
Hawking's\cite{1} demonstration that a non-rotating, uncharged BH of mass $M$
emits thermal radiation at temperature\cite{6}
\begin{equation}
T_{\subsc{BH}}=\frac{\hbar c^3}{8\pi GM k_{\subsc B}}
\label{eq:01-1}
\end{equation}
is mathematically based on quantum field theory in curved spacetime.
This remarkable result is intriguing and beautiful but also a bit subtle and
mysterious.

\begin{figure}[t]
\centerline{\includegraphics[scale=0.3]{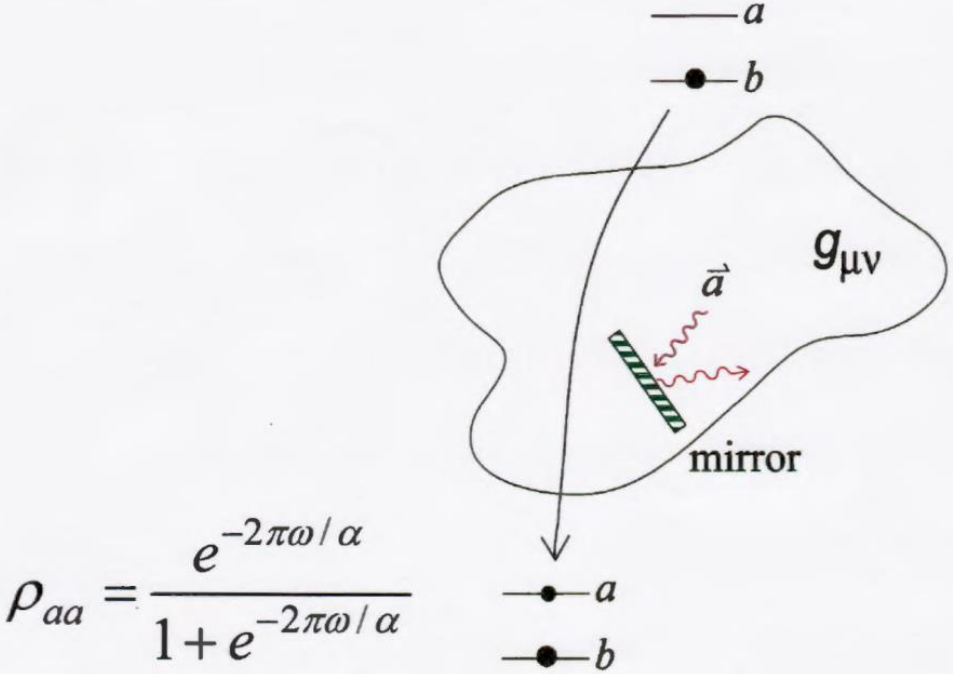}}
\caption{\label{fig:05-3}%
  A detector accelerating through a spacetime region in its field
  vacuum state detects Unruh radiation.
  This happens if in the frame relative-to-which the vacuum modes are defined,
  the atom is accelerating; whether or not the atom is actually accelerating.
  This could even happen if the atom is inertial, and the metric is flat, or
  if there are mirrors modifying the boundary conditions of the spacetime
  modes.}
\end{figure}

From a different point of view, our group of quantum optics and general
relativity aficionados have teamed up to show\cite{2,3} that atoms freely
falling into a BH with the field in the Boulware vacuum (the state of the
field in which no Hawking radiation is emitted by the black hole) emit
radiation which has a thermal energy spectrum (but has phase correlations
between the energy states making the emitted radiation a pure state rather
than a thermal density matrix) which to a distant observer has aspects that
look like (but also aspects that differ from) Hawking radiation.
We call it Horizon Brightened Acceleration Radiation (HBAR).\cite{2}
It is produced solely by emission from the atom while outside the BH.
This work was inspired by quantum optics in flat spacetime, which predicts
that atoms moving with a uniform acceleration emit thermal radiation with
Unruh\cite{4} temperature. 
Although freely falling (having geodesic motion), the atoms seem to a distant
observer to be accelerating in their fall into the black hole, and thus seem
to that observer to be accelerated detectors in the Boulware vacuum (which for
a distant observer is one with no particles). 

However, rather than being excited as though in a thermal bath, they emit
radiation whose energy spectrum as seen by the distant observer looks thermal
with a temperature $T_{\subsc U}$ proportional to their acceleration $\alpha$,
\begin{equation}
T_{\subsc U}=\frac{\hbar \alpha}{2\pi ck_{\subsc B}}\,.
\label{eq:01-2}
\end{equation}
As is explained in the following section, this ``acceleration radiation''
arises from processes in which the atom jumps from the ground state to an
excited state, together with the emission of a photon.\cite{5,scully06}
In quantum optics, such processes are usually discarded because they violate
conservation of energy, and the virtual photons must be quickly reabsorbed in
order to maintain the overall energy conservation.
However, if the atom is accelerated away from the original point of virtual
emission, there is a small probability that the virtual photon will ``get
away'' before it is re-absorbed. 
Alternatively, the Doppler shift of the accelerated atom takes the
otherwise re-absorbed photon out of the atom's bandwidth.
Atom acceleration converts virtual photons into real ones at the expense of the
energy supplied by the external force field driving the center-of-mass motion
of the atom (in Unruh's original case, the acceleration results from an external
force, while in our case, the seeming acceleration is due to gravity).
In an alternate point of view, one can trace the excitation of the atom to a
vacuum fluctuation, which in the usual case is canceled by a succeeding,
correlated fluctuation. 
However, in the accelerated case, the velocity of the atom is different by the
time that correlated fluctuation hits it, giving a Doppler shift which now
means that the fluctuation has the wrong frequency for de-exciting the atom. 

\newpage

Near the event horizon, at radii close to $r_{\subsc g} = 2MG/c^2$, the
Schwarzschild metric is well-approximated by the constant-acceleration Rindler
metric,\cite{Rind60} in which an atom would have a gravitational acceleration of
$\alpha=c^2/2r_{\subsc g}$ (even though to itself it has zero acceleration).
The vacuum state through which it falls is one in which observers at rest
in that frame see no particles.
While in the usual Unruh effect, the atom is excited, in this case, the atom
emits photons whose energy spectrum (as seen by distant stationary observers)
appears to be thermal. 
As a result, the temperature, the HBAR temperature, can be obtained from the
Unruh temperature by plugging $\alpha=c^2/2r_{\subsc g}$ into
Eq.~\eqref{eq:01-2} to find 
\begin{equation}
T_{\subsc{HBAR}}=\frac{\hbar}{2\pi ck_{\subsc B}}\frac{c^2}{ 2r_{\subsc g}}
=T_{\subsc{BH}}\,.
\end{equation}
$T_{\subsc{HBAR}}$ is equal to the temperature of Hawking radiation
\eqref{eq:01-1}. 

This radiation differs from Hawking radiation in that, although the
probability of emission of the various possible energies is proportional to a
thermal spectrum, the emission from any one atom is a pure state, with
definite phase relations between the energies.
Of course if one has many atoms with incoherent times of fall into the black
hole, or if one took into account the recoil of the atom, some of that phase
coherence could be destroyed, making the emission look closer to Hawking
radiation. 

However, the physics is very different from that of the Hawking effect.
Here we have radiation coming from the atoms, whereas Hawking radiation
requires no extra matter (e.g., atoms) and arises just from the BH geometry. 

There are several features of this finding that some have found surprising.
For example one objection could be that the atom is freely falling with proper
acceleration of zero.
Where then does the radiation come from?
However this neglects that the state of the field is assumed to be the
Boulware vacuum state in which the particle content near infinity is zero, but
near the horizon is full of particles (the energy density actually diverges at
the horizon). 
It is those particles that the atom is interacting with.
And from far away, the atom looks as though it is accelerated as it falls into
the black hole.

\enlargethispage{1.0\baselineskip}

In the following section (Sec.~II), we first follow a quantum optics path to
Unruh radiation and compare it to the more usual treatment based on quantum
fields in curved spacetime. 
In Sec.~III, we use two scenarios where, surprisingly, acceleration radiation
is emitted by inertial detectors, for discussing the equivalence principle of
Einstein (in one case, we have a stationary atom interacting with a moving
mirror, and in the other case, we have an atom freely-falling into a black
hole). 
In Sec.~IV, we discuss how Unruh radiation occurs because of the difference
between mode definitions in different frames --- a point of view in which it
is not surprising that an inertial observer would detect acceleration radiation.
In Sec.~V, we present a KMS-inspired method for obtaining the Unruh
temperature, an approach pioneered by Christensen and Duff.\cite{duff}
There, we use the KMS periodicity approach to get the Unruh temperature from
both a field and an atom perspective. 
We summarize in Sec.~VI.

\begin{figure}[t]
\centerline{\includegraphics[scale=0.78]{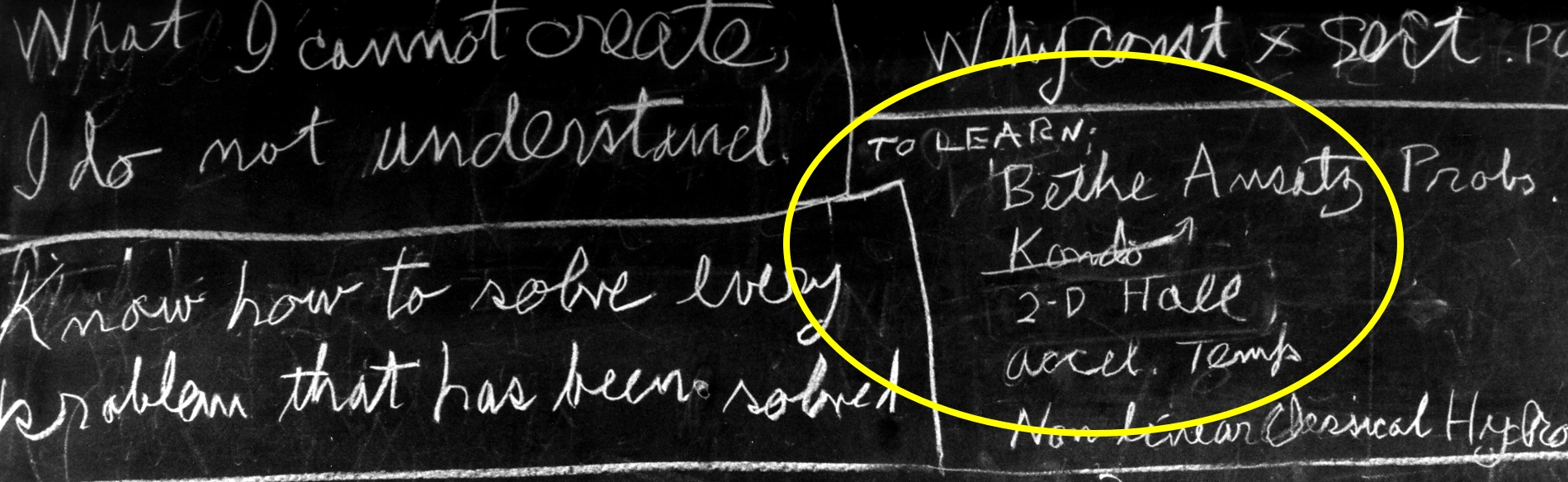}}
\caption{\label{fig:06-5}%
Feynman's blackboard as he left it.
On the bottom-right corner he inscribed ``accel.\ Temp'' under ``TO LEARN:''.
This is a strong indication of the subtlety and interest in this problem
(courtesy of the Archives, California Institute of Technology).} 
\end{figure}

\section*{II. Quantum Optics Route to Obtaining Unruh Radiation in Minkowski
  Coordinates} 
In this section, we provide a simple first principles calculation of the
radiation emitted by an accelerating atom.
This calculation bears similarities to that of Unruh and Wald.\cite{UnruhWald84}
It answers, in part, the implied question of Feynman and Milonni, as in
Fig.~\ref{fig:06-5}. 

Milonni wrote:
\begin{quote}
[A] uniformly accelerated detector [i.e., atom] in the vacuum responds as it
would if it were at rest in a thermal bath at temperature
$T=\hbar a/2\pi ck_{\subsc B}$. 
\emph{It is hardly obvious why this should be} [emphasis added] ---
it took half a century after the birth of the quantum theory of radiation
for the thermal effect of uniform acceleration to be discovered.
\end{quote}

\subsection*{IIa. Accelerating atom in a vacuum}
We consider a two-level atom ($a$ is the excited level and $b$ is the ground
state) with transition frequency $\omega $ moving along the $z$-axis 
in a $1+1$-dimensional spacetime with a uniform acceleration $\alpha$.
The atom trajectory is given by 
\begin{equation}
ct(\bar t)=\ell\sinh\left( \frac{c\bar t}\ell \right)\,,
\quad
z(\bar t)=\ell\cosh\left( \frac{c\bar t}\ell \right)\,,
\label{u1}
\end{equation}
where $t$ is the lab time and $\bar t $ is the proper time for the accelerated
atom,\cite{Rind56} and where
\begin{equation}
\ell=c^2/\alpha
\label{eq:17-33b}
\end{equation}
is the length-scale in the problem.
The interaction Hamiltonian between the atom and an outward-propagating photon
with wave number $k$ reads 
\begin{equation}
\hat{V}(\bar t )=\hbar g\left[ \hat{a}_{k}e^{-i\nu t(\tau )+ikz(\tau )}+%
\text{H.a.}\right] \left( \hat{\sigma}e^{-i\omega \bar t }+\text{H.a.}\right) \,,
\label{u2}
\end{equation}
where operator $\hat{a}_{k}$ is the photon annihilation operator,
$\hat{\sigma}$ is the atomic lowering operator, and $g$ is the atom-field
coupling constant which depends on the atomic dipole moment and on the
electric field in the frame of the atom.

Initially the atom is in the ground state and there are no photons.
If the interaction is weak enough, the state vector of the  atom-field system
at the atomic proper time $\bar t $ can be found using first-order
time-dependent perturbation theory,
\begin{equation}
  \left\vert \psi (\bar t )\right\rangle
  =\left\vert \psi (\tau_{0})\right\rangle -\frac{i}{\hbar }
  \int\limits_{\bar t _{0}}^{\tau }\D\bar t^{\prime }\,
  \hat{V}(\bar t ^{\prime })\left\vert \psi (\tau _{0})\right\rangle \,.
\end{equation}
The probability of excitation of the atom (frequency $\omega $) with
simultaneous emission of a photon with frequency $\nu$ is due to a
counter-rotating term $\hat{a}_{k}^{+}\hat{\sigma}^{+}$ in the interaction
Hamiltonian. 

The probability of this event is
\begin{equation}
P=\frac{1}{\hbar ^{2}}\left\vert \int\limits_{-\infty }^{\infty }
  \D\bar t^{\prime }\,
  \left\langle 1_{k},a\right\vert \hat{V}(\bar t ^{\prime })
  \left\vert 0,b\right\rangle \right\vert ^{2}
=g^{2}\left\vert \int\limits_{-\infty }^{\infty}\D\bar t ^{\prime }\,
  e^{i\nu t(\bar t ^{\prime })-ikz(\tau ^{\prime })}
  e^{i\omega \tau ^{\prime}}\right\vert ^{2}\,,
\end{equation}
where  $\left\vert b\right\rangle$ and $\left\vert a\right\rangle$ are the
ground and excited state of the atom respectively, and $t(\bar t')$ and
$z(\tau')$ are obtained from Eqs.~(\ref{u1}), and using that $k=\nu /c$
and changing the variable of integration to
$x=\frac{\nu \ell}c e^{-c\bar t'/\ell}$, and taking into account that
\begin{equation*}
\int\limits_{0}^{\infty }\D x \,
e^{-ix}x^{ -i\frac{\omega\ell }c - 1 }
=e^{ -\frac12 \frac{\pi\omega\ell }c }\Gamma
\left( -\frac{i\omega\ell}c \right) \,,
\end{equation*}
where $\Gamma(x)$ is the gamma function, and the property
$|\Gamma(-ix)|^{2}=\pi /[x\sinh (\pi x)]$, we finally obtain that the
probability is 
\begin{equation}
P=\frac{2\pi c g^2}{\alpha\omega }
\frac{1}{\exp\left(2\pi \frac{\omega \ell}c\right)-1}\,.
\label{eq:06-1}
\end{equation}

We find that $P$ is proportional to the Planck factor
$1/\left[\exp\left(\frac{2\pi\omega c}{\alpha}\right)-1\right]$
which is the probability that the atom is excited and a photon is emitted.
The Planck factor corresponds to excitation probability with a temperature 
that is proportional to the acceleration $\alpha$,
\begin{equation*}
T_{\subsc U}=\frac{\hbar \alpha}{2\pi c k_{\subsc B}}\,.
\end{equation*}
This can be understood as was discussed in the previous section, as generating
a photon by breaking adiabaticity due to the acceleration of the atom.
Another physical picture involved the promotion of vacuum fluctuations.
In any case, the operator product
$\hat \sigma^\dagger(\bar t) \hat a^\dagger_k(t,z)$
tells us that the (Minkowski) photon is emitted and the atom is excited.

\subsection*{IIb. Excitation of a Static Atom by the Rindler Vacuum}
\begin{figure}[t]
\centerline{\includegraphics[scale=0.5]{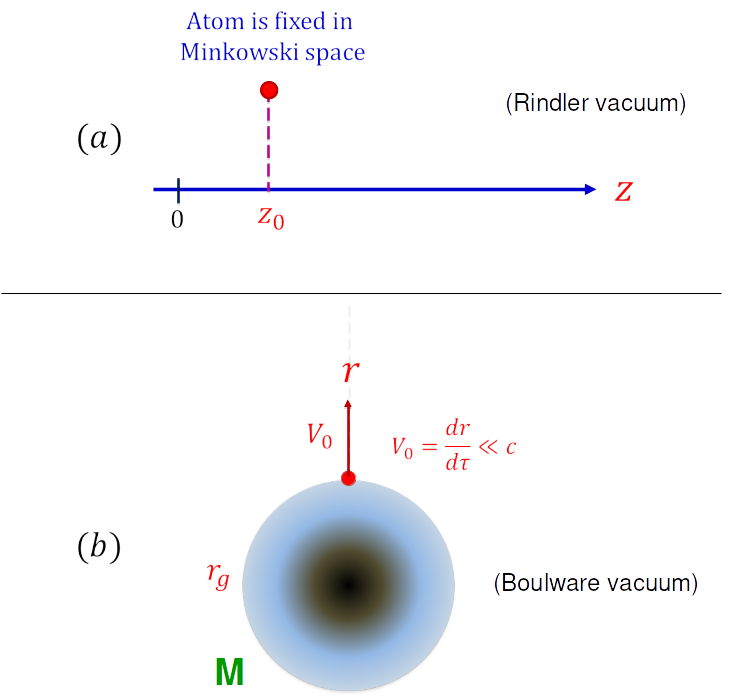}}
\caption{\label{fig:06-4x}%
(a) An atom is fixed in Minkowski spacetime at coordinate $z_{0}$
and the field is in the Rindler vacuum (created by an accelerated mirror).
(b) An atom moves in the vicinity of the BH event horizon in the Boulware
vacuum, emitting acceleration radiation.
These two cases are equivalent to each other, given that the acceleration of
the mirror is related to the BH mass by Eq.~\eqref{eq:09-36}.}
\end{figure}

Having seen that an atom accelerating through the Minkowski vacuum emits
(Minkowski) photons, we consider the ``inverse'' problem of a stationary atom
in an accelerating Rindler vacuum.
To put this in perspective, Sec.~IIa represents the Cavity QED problem of an
atom passing through a stationary cavity.
In this section (IIb), we are essentially dealing with an accelerating
mirror\cite{Moore70}
(with the state of the field being a Rindler-like vacuum) and stationary atom,
as in Fig.~\ref{fig:06-4x}b.
This is the physics behind the present Rindler coordinate analysis.

\begin{figure}[t]
\centerline{\includegraphics[scale=0.6]{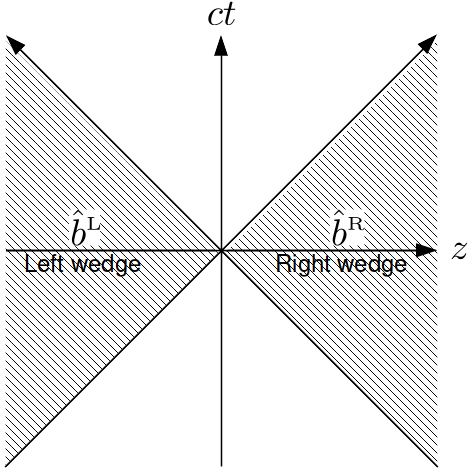}}
\caption{\label{fig:rindwedge}%
Minkowski space divided into four wedges.
Of particular relevance are the right and left wedges, which are called
``the right Rindler wedge'' and ``the left Rindler wedge,'' respectively.}
\end{figure}

We proceed by assuming that an atom is fixed in the inertial reference frame
$(t,z)$ at position $z=z_{0}$ (see Fig.~\ref{fig:06-4x}a).
We make a coordinate transformation into a uniformly accelerating reference
frame, 
\begin{equation}
ct =\ell e^{\bar z/\ell} \sinh\left( \frac{c\bar t}\ell \right)
\,,\quad
z=\ell e^{\bar z/\ell}\cosh\left( \frac{c\bar t}\ell \right)\,,
\label{z2a}
\end{equation}
where $\ell$ is defined in the same way as in Eq.~\eqref{u1}, which gives that
the proper acceleration at $\bar z=0$ is $\alpha$. 
See Fig.~\ref{fig:rindwedge}.

The coordinate transformation (\ref{z2a}) covers only the part of the
Minkowski spacetime with $z>c|t|$ (right Rindler wedge).
It converts the Minkowski spacetime line element
$\D s^{2}=c^{2} \D t^{2}- \D z^{2}$
to the Rindler line element,\cite{Rind60,EinsteinRosenFootnote}
\begin{equation}
\D s^{2}=e^{2a\bar{z}/c^{2}}\left( c^{2}\D \bar{t}^{2}-\D \bar{z}^{2}\right) \,.
\label{Ra}
\end{equation}%
An observer moving along the trajectory $\bar{z}=0$ in the Rindler space is
uniformly-accelerating in the Minkowski space along the trajectory \eqref{u1}, 
which is a special case ($\bar z=0$) of Eq.~\eqref{z2a}.
Normal modes of scalar photons in the conformal metric (\ref{Ra}) take the
same form as the usual positive frequency normal modes in the Minkowski
metric, e.g., one can take them as traveling waves,
\begin{equation}
\phi _{\nu }(\bar{t},\bar{z})=\frac1{\sqrt \nu}e^{-i\nu \bar{t}+ik\bar{z}}\,,
\label{mR}
\end{equation}%
where $\nu $ is the photon angular frequency in the reference frame of the
Rindler space and $k=\pm \nu /c$.
However, the modes (\ref{mR}) are a mixture of positive and negative frequency
modes with respect to the physical Minkowski spacetime.
Therefore, the vacuum state of these modes is not the Minkowski vacuum but
rather the Rindler vacuum, which is what we assume for those modes.

From Eq.~(\ref{z2a}) we obtain $\bar{t}$ and $\bar{z}$ in terms of $t$ and $z$,
\begin{equation}
c\bar{t}(t,z)=\frac{\ell}{2}\ln \left( \frac{z+ct}{z-ct}\right)
\,,\quad
\bar{z}(t,z)=\frac{\ell}{2}\ln\left[\frac{ z^{2}-c^{2}t^{2} }{\ell^2}\right]\,.
\label{z4}
\end{equation}%
The atomic trajectory is obtained from Eq.~(\ref{z4}) by setting the Minkowski
space position to $z=z_{0}$. 
In the Rindler space, the atomic velocity is
\begin{equation}
\bar{V}=\Deriv{\bar z}{\bar t}{}=-\frac{c^{2}t}{z_{0}}\,.
\end{equation}

From the perspective of the atom, it passes through the right Rindler wedge
within the proper time interval  
\begin{equation*}
-\frac{z_{0}}{c}<t<\frac{z_{0}}{c}
\end{equation*}%
for which the atom velocity in the Rindler space $\bar{V}$ changes from $c$
to $-c$.
During this time the atom interacts with the mode (\ref{mR}).
The probability $P$ that the static atom gets excited and a photon in the mode
(\ref{mR}) is generated is given by the integral 
\begin{equation}
P=g^{2}\left\vert \int_{-\frac{z_{0}}{c}}^{\frac{z_{0}}{c}}\D t \,
\phi _{\nu}^{\ast }(t,z_{0})e^{i\omega t}\right\vert ^{2}\,,  \label{g0}
\end{equation}
where $t$ is the proper time for the atom, and $z$ is taken at the atomic
position $z_{0}$.
Using Eqs.~(\ref{mR}) and (\ref{z4}), we obtain (assuming $k=\nu /c$)
\begin{equation}
P=g^2\left\vert
   \int\limits_{-\frac{z_{0}}{c}}^{\frac{z_{0}}{c}}\D t \,
	e^{-i\frac{\nu \ell}c\ln \left[(z_{0}-ct)/\ell\right] 
    +i\omega t}\right\vert ^{2}\,.  \label{p1}
\end{equation}
Changing the integration variable to $x=\omega (z_{0}/c-t)$, we have
\begin{equation}
P=
\frac{g^{2}}{\omega ^{2}}\left\vert \int\limits_{0}^{
    \frac{2\omega z_{0}}{c}}\D x \, e^{ix}x^{ i\frac{\nu \ell}c }
\right\vert ^{2}\,.
\label{eq:09-24}
\end{equation}
Using that
\begin{equation*}
\int\limits_{0}^{\frac{2\omega z_{0}}{c}}\D x \, e^{ix}
x^{ i \frac{ic\ell }c }=e^{ -\frac\pi2 \frac{\nu\ell }c }
\gamma\left(1 + i\frac{\nu\ell }c,-i\frac{2\omega z_0}c\right)\,,
\end{equation*}
where $\gamma (s,x)$ is the incomplete lower gamma function
which has the asymptotic behavior
$\gamma (s,-ix)\rightarrow i\Gamma (s),\text{ as } x\rightarrow\infty$,
we find that the probability in Eq.\ \eqref{eq:09-24} is
\begin{equation}
P=\frac{g^{2}}{\omega ^{2}}e^{ -\pi \frac{\nu\ell }c }
\left\vert\gamma\left(1+i\frac{\nu\ell }c,-i\frac{2\omega z_{0}}{c}
\right) \right\vert ^{2}\,.
\label{t5}
\end{equation}
In the limit $z_{0}\gg c/\omega $ we have
\begin{align}
&\left\vert\gamma\left(1+i\frac{\nu\ell }c,-i\frac{2\omega z_{0}}{c}\right)
\right\vert ^{2}\approx\left\vert\Gamma\left(1+i\frac{\nu\ell }c
\right)\right\vert ^{2}
=
\left(\frac{\nu\ell }c\right)^{2}
\left\vert\Gamma\left(i\frac{\nu\ell }c\right)\right\vert ^{2}
=\frac{\nu\ell }c\frac\pi{\sinh\left(\pi \frac{\nu\ell }c\right)}
\end{align}
which yields that the probability for exciting the atom along with emission of
a $\nu$-photon is 
\begin{equation}
  P\approx \frac{2\pi \nu\ell g^2}{c\omega^2}
  \frac{1}{\exp\left[2\pi\frac{\nu\ell}c\right]-1}\,.
\label{P0}
\end{equation}
Notice that in our present case of a stationary atom  in the Rindler vacuum
the Planck factor is $\nu$-dependent, whereas in the case of the accelerating
atom, the Planck factor in the analogous Eq.~\eqref{eq:06-1} in Sec.~II
is $\omega$-dependent.
It is the emitted radiation by the stationary atom which is thermal, not the
excitation of the atom. 

The probability of photon absorption is obtained by changing
$\nu\rightarrow -\nu $.
Equation~(\ref{P0}) yields 
\begin{equation}
P_{\subsc{abs}}=\exp\left[ \Big.	2\pi \frac{ \nu\ell}c\right]
P_{\subsc{exc}}\,.
\label{eq:11-27}
\end{equation}%
However, if we use the more accurate Eq.~(\ref{t5}), we obtain
\begin{equation}
P_{\subsc{abs}}
=\exp\left[\Big.2\pi \frac{ \nu\ell}c\right]
\frac{\left\vert\gamma\left(1+i\frac{\nu\ell}c,
      i\frac{2\omega z_{0}}{c}\right)\right\vert^2}
     {\left\vert \gamma\left(1+i\frac{\nu\ell}c
     ,-i\frac{2\omega z_{0}}{c}\right)\right\vert^2}P_{\subsc{exc}}\,,
\label{eq:11-28}
\end{equation}
which is thermal only in the limit $z_0 \gg c/\omega$.

\section*{III. Acceleration radiation and the equivalence principle using %
     Unruh--Minkowski modes} 
Let us approach the question of the relation between accelerated motion of
either the mirror or the atom in an accelerated vacuum in a different way.

\begin{figure}[t]
\centerline{\includegraphics[scale=0.3]{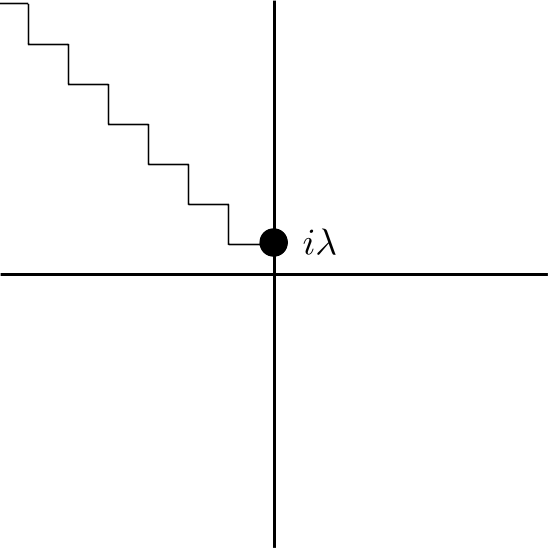}
\quad
\includegraphics[scale=0.3]{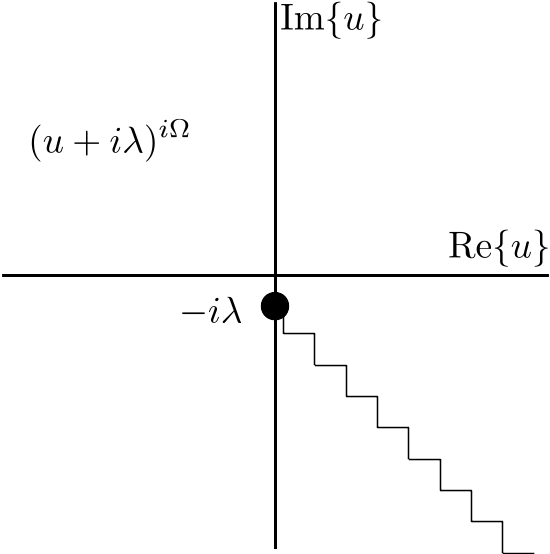}}
\caption{\label{fig:15-8}%
  The definition of the branch cut in Eq.~\eqref{eq:02-u1} for positive
  $\lambda$.} 
\end{figure}

Consider the function 
\begin{equation}
f(t)=\lim_{\lambda\rightarrow0^+}
\left(\frac{t \pm i\lambda}{\ell/c}\right) ^{i\Omega}
\label{eq:02-u1}
\end{equation}
where 
$\Omega=\nu'\ell/c$ is some dimensionless frequency,
and $\lambda\rightarrow 0_+$.
This $i\lambda$ prescription is to indicate the sector of the complex $t$
plane in which we place the branch-cut of the function.
I.e., in both cases, we take $\lambda\rightarrow 0^+$, but 
$-i\lambda$ indicates that the branch cut is in the upper-half complex $t$-plane,
while $+i\lambda$ would indicate that is in the lower-half complex $t$-plane.
See Fig.~\ref{fig:15-8}.

To determine the frequency content of $f(t)$ in Eq.\ \eqref{eq:02-u1}, we
consider the integral
\begin{equation}
F(\omega)=\int_{-\infty}^\infty\D t \,e^{i\omega t}
\left(\Big.\frac{t-i\lambda}{\ell/c}\right)^{i\Omega}.
\end{equation}
If $\omega<0$, the integral can be completed in the lower-half complex
$t$-plane, giving $F(\omega)=0$ for all values of $\Omega$.
Thus, the Fourier transform of $f(t)$ is non-zero only for positive $\omega$,
i.e., it is a purely positive-frequency function. 

Similarly, $[c(t+i\lambda)/\ell]^{i\Omega}$ is a purely negative-frequency
function for all values of $\Omega$.
These functions thus form a complete set of functions over $t$ under the
Klein--Gordon inner-product 
\begin{equation}
\finner{f}{g}=-\frac{i}{2}
\int \D t \,\left( \Big.f \pderiv t{} g^*
   - g^* \pderiv t{} f\right)\,.
\label{eq:01-7}
\end{equation}

We will use a complete set of modes similar to Eq.~\eqref{eq:02-u1} to examine
two different situations: the motion of a mirror by a stationary atom,
and the motion of a two-level atom (or detector) in the presence of a mirror,
both interacting with a massless scalar field $\hat\Phi$.
We will work in 1+1 dimensional spacetime.
These will be special cases of systems which some of us have examined
previously.\cite{2,3,LPP} 

In the first case, we will have a mirror at rest in the ordinary Minkowski
vacuum state, i.e., the state in which one would ordinarily say there are no
particle excitations of the scalar field.
The atom, however, is accelerated with constant acceleration $\alpha$,
following the trajectory in Eq.~\eqref{u1}.
We specialize to the case where the atom's closest approach to the mirror is
given by the distance $\ell$, defined in Eq.\ \eqref{eq:17-33b}.
See Fig.~\ref{fig:15-9}.
In the second case, we swap the behavior of the atom and the mirror,
and choose a different initial state for the field.
The mirror will follow a trajectory of constant acceleration, Eq.~\eqref{u1},
while the atom will be at rest.
Again, the distance of closest approach of the mirror to the atom would
be~$\ell$. 
In this case, we will take the state of the quantum field to be the so-called
Rindler vacuum.
This is the state in which the accelerated mirror sees the quantum field
as containing no particles.
See Fig.~\ref{fig:15-10}.
It is in some sense an approximate weak equivalence principle%
\footnote{This is of course only a crude approximation for the weak
  equivalence principle, since when one is in a bumper car that decelerates
  rapidly when it hits another one against a rail that prevents it from
  accelerating, one will feel different from when one is in a bumper car
  against a rail that does not accelerate when another hits it, even though
  the relative acceleration is the same in the two cases.
  See Fig.\ \ref{fig:06-4}, where it is seen that the full equivalence
  principle is between an accelerating mirror (B) and an atom freely-falling
  into a black hole (C).
  There, we find that the spectra are equivalent.
  However, while the spectra of the accelerating atom (A) and the accelerating
  mirror (B) are strikingly similar they are different, and therefore, the two
  cases are not equivalent.}
analog of the first case.\cite{fullwilson}
In both cases, the mirror sees no photons, and the mirror and atom have the
same relative accelerations with respect to each other. 

In each case, we look at the interaction between the atom and the field only
to lowest order in the coupling constant. 

Let us look at the flat spacetime examples first and take the coordinates
$(t,z)$ to be the usual Minkowski coordinates such that the metric is 
\begin{equation}
\D s^2= \D t^2-\D z^2\,.
\end{equation}
Let us define dimensionless null coordinates $u$ and $v$,
\begin{equation}
u = (ct-z)/\ell\,, \quad
v = (ct+z)/\ell\,.
\label{eq:01-u5}
\end{equation}
The equation of motion for a massless scalar field is
(the massless Klein--Gordon equation)
\begin{equation}
\frac1{c^2} \partial_t^2 \phi -\partial_z^2 \phi=0\,,
\label{eq:02-u6}
\end{equation}
which, in terms of the null coordinates $u$ and $v$ in
Eq.~\eqref{eq:01-u5} is
\begin{equation}
\partial_u\partial_v\phi=0\,,
\label{eq:02-u7}
\end{equation}
where we use the notation
$\partial_\xi \equiv \pderiv\xi{}$.

The plane-wave modes of the field, which are commonly used for expanding
solutions of Eqs.~\eqref{eq:02-u6} or \eqref{eq:02-u7}, are
\begin{align}
\phi_{\omega +}= \frac1{ \sqrt{4\pi\abs\omega} }
e^{-i\omega(ct-z)/c}= \frac1{ \sqrt{4\pi\abs\omega} }
e^{i\omega\ell/c \ u}\,,
\label{eq:20-38a}
\\
\phi_{\omega -}= \frac1{ \sqrt{4\pi\abs\omega} }e^{-i\omega(ct+z)/c}
=\frac1{ \sqrt{4\pi\abs\omega} }e^{i\omega\ell/c \ v}\,,
\label{eq:20-38b}
\end{align}
where $\omega\pm$ correspond to right- and left-moving solutions, respectively.

In terms of the Klein--Gordon norm for the fields, Eq.~\eqref{eq:01-7},
the modes with $\omega>0$ have a positive value for the norm, while those for
$\omega<0$ have a negative norm. 
We however, use a different complete set of modes, Eq.~\eqref{eq:17-33a} below,
which are similar to Eq.~\eqref{eq:02-u1}, for expanding solutions of
Eq.~\eqref{eq:02-u7}. 

Instead of the solutions \eqref{eq:20-38a} and \eqref{eq:20-38b}, we elect to
use a complete set of modes for the field by
\begin{equation}
\hat\phi_{\Omega+}= 
\frac{e^{-\pi\Omega/2} }{ \sqrt{8\pi\Omega\sinh(\pi\Omega)} }
\lim_{\lambda\rightarrow0^+}(u-i\lambda)^{i\Omega}\,,
\label{eq:17-33a}
\end{equation}
where we normalized Eq.~\eqref{eq:02-u1} and use a different variable, $u$.
These are a complete set of positive norm (often called the positive frequency
Unruh--Minkowski modes,\cite{4,UnruhWald84})
even though $\Omega$ takes all values positive and negative. 
The negative-norm modes are just the complex-conjugate of these (due to the
sign of $i\lambda$, or ultimately, the definition of the branch-cut).

\subsection*{IIIa. Accelerating atom}
We are now going to place a mirror at position $z=0$.
We will take the boundary conditions on the solutions $\phi$
that they be zero at the mirror.
The solutions of Eq.~\eqref{eq:02-u7} then are of the form
\begin{equation}
\phi(u,v)= g(u)-g(v)\,,
\label{eq:17-34}
\end{equation}
for some function $g$.
Since at $z=0$, the null coordinates are both $u=v=ct/\ell$, then we see that
$\phi(z=0,t)=g(t)-g(t)=0$, whish satisfies the boundary conditions.
Using Eq.~\eqref{eq:17-34} and the modes \eqref{eq:17-33a}, we have that the
modes satisfying the boundary conditions are
\begin{equation}
\phi_\Omega(u,v)=
\frac{ e^{-\pi\Omega/2} }{ \sqrt{4\Omega \sinh(\pi\Omega)} }
\lim_{\lambda\rightarrow0^+}\left[\Big.(u-i\lambda)^{i\Omega}
	-(v-i\lambda)^{i\Omega}\right]\,.
\label{eq:21-b35}
\end{equation}

For the two-level atom, let us define the two states $\ket{b}$ as the ground
state of the atom and $\ket{a}$ as the excited state, with proper energy
$\omega$, and the atomic raising operator $\hat\sigma^\dagger$, which takes
$\sigma^\dagger\ket{b}=\ket{a}$, having time dependence $e^{i\omega\bar t}$
in the interaction picture, where $\bar t$ is the proper time of the atom. 

\begin{figure}[t]
\centerline{\includegraphics[scale=0.4]{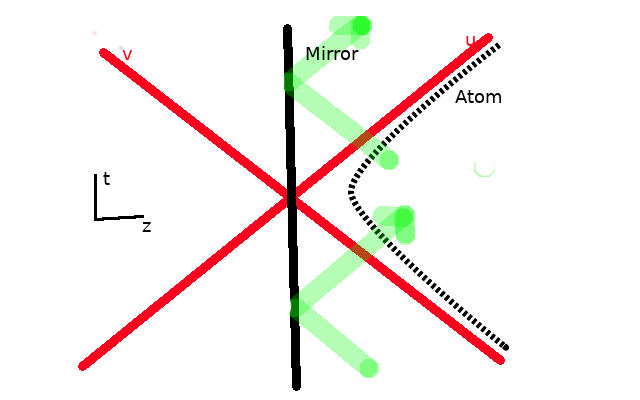}}
\caption{\label{fig:15-9}%
An atom accelerates in the presence of a stationary mirror.
Initially, the atom is in its ground state $\ket b$,
and the field is in the Minkowski vacuum $\ket{ 0_{\subsc M} }$.
There is some amplitude $\ket{ \mathcal A_{ \subsc{ex} } }$
for the atom to become excited.
We note that since the mirror destroys the Lorentz invariance
(in contrast to the true, Lorentz-invariant vacuum case in Sec.~V),
the state seen by the accelerated atom is not a static thermal bath,
but this aspect does not matter for our conclusions.}
\end{figure}

We can write the quantum field $\hat\Phi$ in terms of the null coordinates $u$
and $v$ 
\begin{equation}
\hat\Phi(u,v)=\int \D\Omega \,
\left( \Big.{\hat a}_\Omega\phi_\Omega(u,v)+
	{\hat a}^\dagger_{\Omega}\phi_\Omega^*(u,v)\right)\,.
\end{equation}
In terms of the null coordinates \eqref{eq:01-u5}, the path of the particle
\eqref{u1} is
\begin{equation}
u(\bar t) =- e^{-c\bar t/\ell}\,,\quad
v(\bar t) =  e^{ c\bar t/\ell}\,.
\end{equation}
The interaction between the atom and the field will be taken to be
\begin{equation}
\hat H_{\subsc I}=g\left( \Big.\hat\sigma e^{-i\omega \bar t} +
\hat\sigma^\dagger e^{i\omega\bar t}\right)
w^\mu \partial_\mu \hat\Phi\,,
\label{eq:26-37}
\end{equation}
where $w^\mu$ is the four velocity of the atom, and $\bar t$ is the proper
time along the path of the detector.
In the frame of the atom, it is stationary, thus we have
\begin{equation}
w^\mu \partial_\mu \hat\Phi=\partial_{\bar t} \hat\Phi\,,
\end{equation}
where the derivative is evaluated along the path of the the atom. 
This interaction is chosen because it makes the field $\hat\Phi$ an
ohmic-coupled  bath for the detector, in the nomenclature of Caldera and
Leggett.\cite{CL}
See Fig.~\ref{fig:15-9}.

Since the atom begins in its ground state, and the quantum field in the
Minkowski vacuum state, in the atom-field interaction, the only term that
contributes to the probability amplitude that the atoms becomes excited is the 
``counter-rotating'' term, in the language of quantum optics.
I.e., we need terms that look like $\hat \sigma^\dagger \hat a^\dagger$.
If the atom is not accelerated, such counter-rotating terms will give zero
when integrated over time.
However, using the above definition of the field, and the fact that the
time-dependence of the atomic raising operator $\hat\sigma^\dagger$ is
$e^{i\omega \bar t}$, we get an excitation amplitude of
\begin{align}
\ket{\mathcal A_{ \subsc {ex} }}
&=\bra{a}g\int \D \bar t \,\left[ \Big.
	\hat\sigma^\dagger e^{i\hat\omega\bar t}
	+\hat\sigma e^{-i\hat\omega\bar t}\right]
\partial_{\bar t}\hat \Phi
	\left( \Big.-e^{-c\bar t/\ell}, e^{ c\bar t/\ell}\right)
\ket{ b, 0_{\subsc M} }
\\&=g\int_{-T}^T \D \bar t \,e^{i\omega \bar t}
\nonumber
\\&\qquad\times
\int_{-\infty}^\infty \D \Omega \,
\frac{i\Omega c}{\ell}\left[ \Big.
	(-e^{-c\bar t/\ell})^{-i\Omega}+
	( e^{ c\bar t/\ell})^{-i\Omega}\right]
\frac{ e^{-\pi\Omega/2} }
	{ \sqrt{8\pi\Omega\sinh(\pi\Omega)} }\hat a_\Omega^\dagger 
\ket{0_{\subsc R}}\,,
\label{eq:17-41}
\end{align}
since $\partial_{\bar t}\phi_\Omega^*(u,v)$ is
\begin{equation}
\partial_{\bar t} \phi_\Omega^*(u,v)=
\frac{i\Omega c}\ell\frac{ e^{-\pi\Omega/2} }
	{ \sqrt{8\pi\Omega \sinh(\pi\Omega)} }
\lim_{\lambda\rightarrow0^+}
\left[ \Big.(u+i\lambda)^{-i\Omega}
	+(v+i\lambda)^{-i\Omega}\right]\,,
\end{equation}
where we used Eq.~\eqref{eq:21-b35}.
I.e., the first-order excitation is due to the
$\hat\sigma^\dagger {\hat a}^\dagger$ term,
a product of the counter-rotating terms in the quantum optics nomenclature.
If $\Omega>0$, then the second term in the square brackets will be zero after
integration over $\bar t$, while if $\Omega<0$, it is the first term that will
be zero.
Now $(-x+i\lambda)^{i\Omega}= x^{i\Omega}e^{-\pi\Omega}$ for positive $x$
since one must take the contour around the upper ${-x}$ complex values so that
$(-1)^{i\Omega}=(e^{+i\pi})^{i\Omega}=e^{-\pi\Omega}$.
See Fig.~\ref{fig:15-8}.
The integral in Eq.~\eqref{eq:17-41} thus
becomes (in the limit that $T\rightarrow \infty$) 
\begin{equation}
\ket{{\mathcal A}_{ \subsc{ex} }}
\approx 2T \frac{ e^{-\pi\omega/2a} }
	{ \sqrt{8\pi\omega\sinh(\pi\omega/a)/a} }
\left( \Big.\hat a_{ \omega/a}^\dagger-
	\hat a_{-\omega/a}^\dagger\right)\ket{ 0_{\subsc M} }\,,
\end{equation}
where $\hat a^\dagger_{\omega}$ and $\hat a^\dagger_{-\omega}$
are the creation operators for the right- and left-moving modes,
Eqs.~\eqref{eq:20-38a} and \eqref{eq:20-38b}, respectively.
The probability of atomic excitation is 
\begin{equation}
P_{\subsc {ex} }=
\inner{\mathcal A_{\subsc {ex} }}{\mathcal A_{ \subsc{ex} }}\,,
\end{equation}
which is proportional to the thermal factor $1/(e^{2\pi\omega/a}-1)$.

We note that this is interesting in that there is really no horizon hiding the
partner particles from the quantum field from the detector.
There is entanglement between the incoming field in the right Rindler wedge
and that behind the incoming horizon.
But the latter gets reflected out by the mirror.
Thus the entanglement in the Minkowski vacuum occurs between the ingoing modes
in the right Rindler wedge and the outgoing modes in that same wedge, instead
of being hidden behind the horizon. 

We can ask whether or not the system is truly thermal by comparing the
probability of emission of radiation by an excited accelerated atom with the
absorption of the counter-rotating term by the unexcited atom.

\subsection*{IIIb. Accelerating mirror}
In the second case, we consider an accelerated mirror, with a stationary
detector whose surface is at $uv=-1$, and the field initially in the Rindler
vacuum (as defined by Fulling\cite{Fulling}).
With the mirror accelerated, the field is expanded in terms
of the positive-norm Rindler modes,\cite{Rind60}
\begin{equation}
\begin{aligned}
\bar\phi_{\Omega++}=
\frac1{ \sqrt{4\pi\Omega} }
\begin{cases}
u^{-i\Omega},\ &u>0
\\ 0,\ &u<0
\end{cases}
\qquad \quad
\bar\phi_{\Omega+-}
=
\frac1{ \sqrt{4\pi\Omega} }
\begin{cases}
0,\ &u>0
\\(-u)^{i\Omega},\ &u<0
\end{cases}
\\
\bar\phi_{\Omega-+}=\frac1{ \sqrt{4\pi\Omega} }
\begin{cases}
v^{-i\Omega},\ &v>0\\ 0,\ &u<0
\end{cases}
\qquad \quad
\bar\phi_{\Omega--}=\frac1{ \sqrt{4\pi\Omega} }
\begin{cases}
0,\ &u>0\\(-v)^{i\Omega},\ &v<0
\end{cases}
\label{eq:26-41}
\end{aligned}
\end{equation}
with positive $\Omega$.

\begin{figure}[t]
\centerline{\includegraphics[scale=0.4]{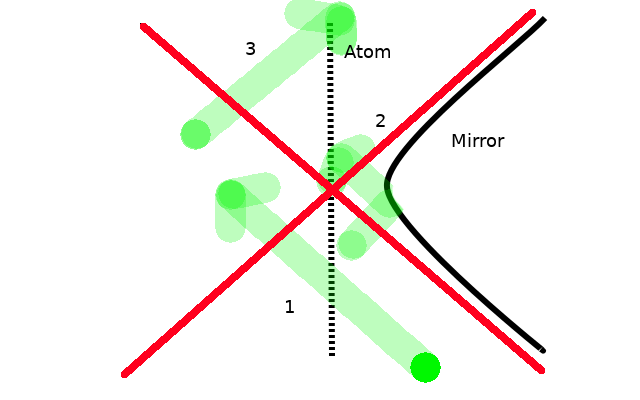}}
\caption{\label{fig:15-10}%
A stationary atom with a moving mirror.
While usually the spatial left- and right-moving modes are independent, in
this scenario, we have three families of modes which interact with the atom.
The first (labeled `2'), consists of right- and left-moving components, with
relative phase $(-1)$ between them, so that they vanish at the mirror.
Also in this spacetime are left-moving modes and right-moving modes (labeled
`1' and `3', respectively) which do not interact with the mirror. 
Those have a random phase relationship to one another.
The mirror follows a trajectory of constant acceleration, and the atom is at
rest. 
The three cases depicted are
(1) the positive-norm ``1-modes'' that originate before the past right null
asymptote of the mirror trajectory and travel to the left; 
(2) the positive-norm ``2-modes'' that originate from the left before the
past extension of the future null asymptote for the mirror, bounce off the
mirror, and continue traveling to the left after the past null asymptote of the
mirror;
(3) the positive-norm ``3-modes'' that originate from the left after the
extension of the future null asymptote of the mirror and travel to the right.
The field is in the Rindler vacuum state, which means that each of the
three types of modes above (and the ``4-modes'' (not depicted) that are to the
right of the mirror and do not interact with the atom) are independent of
(unentangled and uncorrelated with) any of the other modes and have no
particles detectable by accelerated observers in either Rindler wedge. 
The atom is at rest (moving along an inertial static world line) at
distance $\ell$ from the closest approach of the mirror.}
\end{figure}

Because of the mirror, this spacetime features the following modes, which are
superposition of the basic positive-frequency modes, \eqref{eq:26-41}.
See Fig.~\ref{fig:15-10}.
We have the positive-norm ``1-modes,'' which are left-moving modes
in the negative $v$ region (and are zero elsewhere),
\begin{equation}
  \bar\phi_{\Omega 1} = \frac1{ \sqrt{4\pi\Omega} } (-v)^{ i\Omega}
  \,;\quad\quad v<0\,, 
\label{eq:19-35}
\end{equation}
and we have the positive-norm ``3-modes,'' which are right-moving modes
in the positive $u$ region (and zero elsewhere),
\begin{equation}
  \bar\phi_{\Omega 3} = \frac1{ \sqrt{4\pi\Omega} } ( u)^{-i\Omega}
  \,; \quad\quad u>0\,.
\label{eq:19-36}
\end{equation}
In these regions, $v<0$ and $u>0$, there is no mirror.
We have the positive-norm ``2-modes,'' which interact with the mirror.
The region of the spacetime with negative $u$ and positive $v$ contains the
mirror, which lies on the surface $uv=-1$.
These modes are a superposition of the positive-norm
left- and right-moving Rindler modes, Eq.~\eqref{eq:26-41},
which vanish at the mirror.
They are 
\begin{equation}
\bar \phi_{\Omega 2}(u,v)=\frac1{ \sqrt{4\pi\Omega} }
\left[ \Big.	( u)^{ i\Omega}-(-v)^{-i\Omega}\right]\,,
\label{eq:19-37}
\end{equation}
and zero elsewhere.
These are a bit subtly-defined, because the right-moving piece is defined for
$u<0$, but the left-moving part is in $v>0$, see Fig.~\ref{fig:15-10}.
We also have the ``4-modes'' (not shown in the figure).
These are confined to the region $u<0$ and $v>0$ (the right wedge) and vanish
at the mirror, but do not interact with the atom, so we ignore them.

In terms of the positive-norm mode families which interact with the atom,
Eqs.~\eqref{eq:19-35}, \eqref{eq:19-36}, and \eqref{eq:19-37},
the field $\hat\Phi$ is
\begin{equation}
\hat\Phi(u,v) = \sum_{i=1}^3\left(
	\int_0^\infty \D \Omega \;{\hat b}_{\Omega i} \phi_{\Omega i}+
	\mbox{H.a.}\right)\,,
\label{eq:26-48}
\end{equation}
where the summation over $i$ is to include all three mode types.
The atom travels along the path $u=v=ct/\ell$.
The state of the field, the Rindler vacuum, $\ket{0_{\subsc R}}$,
is defined by $\hat b_{\Omega i}\ket{0_{\subsc R}}=0$ for all values of $\Omega$,
and the only terms in the amplitude which survive if the detector is initially
in its ground state $\ket{b}$ are 
\begin{equation}
\ket{ A_{ \subsc{ex} } }=
\int_{-T}^T \D t \int_0^\infty \D\Omega \;
  e^{i\omega t}\partial_t\phi^*_{\Omega i}(t,t){\hat b}^\dagger_{\Omega i}
  \ket{0_{\subsc R}}\,,
\label{eq:26-50}
\end{equation}
where $2T$ is the interaction time, and we used Eq.~\eqref{eq:26-48} for the
field $\hat\Phi$ and Eq.~\eqref{eq:26-37} for the interaction Hamiltonian.

To calculate \eqref{eq:26-50} for infinite interaction time $T$, we first
compute $W_{\Omega\pm}$, where
\begin{equation}
W_{\Omega\pm}=\pm\int_0^{\pm\infty} \D t \,e^{i\omega t} \partial_t
\left( \Big.\pm\frac{ct}\ell\right)^{\pm i\Omega}\,.
\label{eq:21-b53}
\end{equation}
To compute $W_{\Omega+}$, we rotate the contour of integration from the real
$t$-axis to the imaginary $t$-axis, with $t_{\subsc I}=\IM{t}$ and where the
branch-cut is not in the first quadrant of the complex $t$-plane. 
\begin{equation}
W_{\Omega+}=\int_0^\infty \D t_{\subsc I} \,
e^{-\omega t_I} \partial_{ t_{\subsc I} }
\left( \Big.\frac{ ct_{\subsc I} }\ell\right)^{i\Omega}e^{-\pi\Omega/2} \,.
\end{equation}
Changing integration variables from $t_{\subsc I}$ to $x=\omega t_{\subsc I}$
we get 
\begin{align}
W_\Omega&=
\left( \Big.\frac{c}{\omega\ell}\right) ^{i\Omega}
e^{-\pi\Omega/2}\int_0^\infty \D x \,
e^{-x} \partial_x  x^{i\Omega}
=(i\Omega)\left( \Big.\frac{c}{\omega\ell}\right) ^{i\Omega}
e^{-\pi\Omega/2} \,\Gamma(i\Omega)
\notag\\&=
i\frac{ \sqrt{\pi\Omega} \, e^{-\pi\Omega/2} }
      { \sqrt{\sinh(\pi\Omega)} }
e^{i\varphi(\Omega)}
\left(\Big.\frac{c}{\omega\ell}\right) ^{i\Omega}\,,
\label{eq:16-32}
\end{align}
where $\varphi(\Omega)$ is the slowly-varying phase of the complex argument
gamma function $\Gamma(i\Omega)$, which starts at $-\pi/2$ for $\Omega=0$ and
reaches $0$ only once $\Omega \approx 3$, by which time
$e^{-\pi\Omega}\sqrt{ \frac\Omega{2\sinh(\pi\Omega)} }$ will have dropped by a
factor of about $10^3$.
I.e., the phase  of $\Gamma(i\Omega)$ is essentially constant
over the range in which the $\Gamma(i\Omega)$ is non-zero.

Similarly, one can rotate the contour in Eq.~\eqref{eq:21-b53} the other way
and evaluate  
\begin{equation}
-W_{\Omega-}=\int_\infty^0 \D t \,e^{i\omega t}
\partial_t\left( \Big.	i \frac{ct}{\ell}\right)^{i\Omega}
=\left( \Big.\frac{c}{\omega\ell}\right) ^{i\Omega}
\int_\infty^0 \D x \,e^{-\pi\Omega/2} e^{-x}\partial_x x^{i\Omega}\,.
\end{equation}
We thus find that $W_\Omega=W_{\Omega+}=(W_{\Omega-})^*$, and therefore,
the excitation amplitude per $\Omega$ is
\begin{equation}
\ket{{\mathcal A_\Omega}}=\frac g{ \sqrt{4\pi\Omega} }
\left[ \Big.\hat b^\dagger_{\Omega1}W_\Omega^*
+\hat b^\dagger_{\Omega3}W_\Omega+\hat b^\dagger_{\Omega2}
\left( \Big.W_\Omega - W_\Omega^*\right)\right]\,,
\label{eq:17-57}
\end{equation}
where, using \eqref{eq:26-50}, the full amplitude is
\begin{equation}
\ket{ A_{\subsc{ex}} }=\int \D \Omega \,\ket{A_\Omega}\,.
\end{equation}

Integrating the amplitude $\ket{A_\Omega}$ in Eq.~\eqref{eq:17-57}
over $\Omega$ gives some constant which is independent of the
frequency of the atom, and certainly not thermal.
However, the probability of emitting a mode with frequency $\Omega$ is
proportional to a thermal factor 
\begin{equation}
P_\Omega=\vert\bra{ 0_{\subsc R} } {\hat b}_{\Omega i} 
\ket{\mathcal A_\Omega}\vert^2
\propto\frac{1}{ 1 - \exp[2\pi\Omega] }
\end{equation}
which was also found in Ref.~\citenum{3}.

Thus, an accelerated atom above a stationary mirror with the field in the
Minkowski vacuum (no particles detected by the mirror as striking the
stationary mirror) is excited with a probability proportional to the thermal
factor, while an accelerated mirror above a stationary atom, with the field in
the Rindler vacuum (i.e., no particles detected by the mirror as striking the
mirror) emits Rindler modes with a probability proportional to the thermal
factor.
We must distinguish this statement from stating that the atom emits particles
into a thermal state.
The atom emits modes with correlations between the modes, given by the phase
factor $i(a/\omega)^{i\Omega}e^{\varphi(\Omega)}$, as in Eq.~\eqref{eq:16-32}.
I.e., what an unaccelerated atom below the accelerated mirror emits is a pure
state, not a thermal state (a mixed state);
albeit, the probability distribution over Rindler energies is proportional to
a thermal factor.  
There is thus some crude approximate form of the equivalence principle in play
here. 

Hawking showed that a black hole emits thermal radiation.
While an observer at infinity sees the black hole as in some sense stationary,
a static observer or atom near the horizon is accelerated with constant
acceleration. 
The Hartle--Hawking state of the field near the black hole looks like a
thermal state to such a static observer, but looks much more like a vacuum
state to a freely-falling observer.
We can again look at two cases, the one analyzed by Hawking, in which the atom
is accelerated and near the horizon, while the state is the vacuum state as
far as the horizon is concerned (although it is a state in thermal equilibrium
with a temperature inversely-proportional to the mass for an observer far
away).
The second case is where the atom is in free fall into the horizon,
while the state of the field is the so-called Boulware vacuum
(the analog of the Rindler vacuum in the curved spacetime of the
Schwarzschild metric of a non-rotating black hole),
where a distant observer sees nothing coming out of the black hole.

\section*{IV. Acceleration Radiation and the Equivalence Principle}
\begin{figure}[t]
\centerline{\includegraphics[scale=0.7]{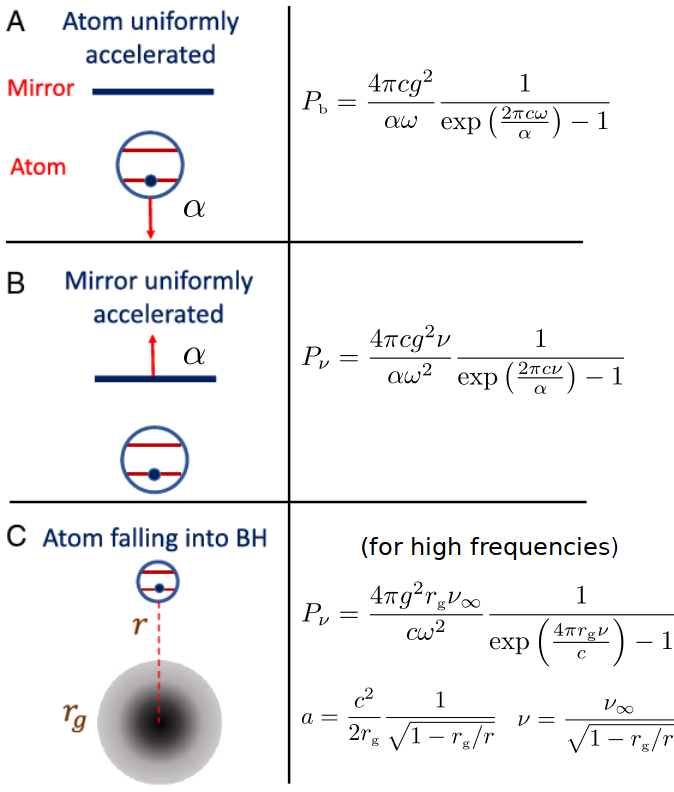}}
\caption{\label{fig:06-4}%
The three physical cases which we consider:
(A) An atom uniformally accelerated in Minkowski space in the presence of a
stationary mirror with the Minkowski vacuum.
(B) A stationary atom in Minkowski space in the presence of an accelerating
mirror and the Rindler vacuum.
(C) An atom in free-fall in the Schwarzschild metric in the Boulware vacuum.
In all three sub-figures, we indicate the probability of atomic excitation
(atomic frequency $\omega$) in the first case or with an excitation probability
at high frequency for the electromagnetic field mode with frequency $\nu$.
Cases (B) and (C) are similar because in both cases, the atom is freely-falling,
but still emits radiation.
Case A is a physically-different case, because the \emph{atom} has non-zero
proper acceleration, and it is the atom that is thermally-excited, giving
physically-different results.} 
\end{figure}

In this section, we discuss acceleration radiation from atoms which do not
accelerate, and show the approximate equivalence between atoms freely-falling
into a Schwarzschild black hole and stationary atoms (in Minkowski space) in
the presence of an accelerating mirror. 
See Fig.~\ref{fig:06-4}.

\begin{figure}[t]
\centerline{\includegraphics[scale=0.5]{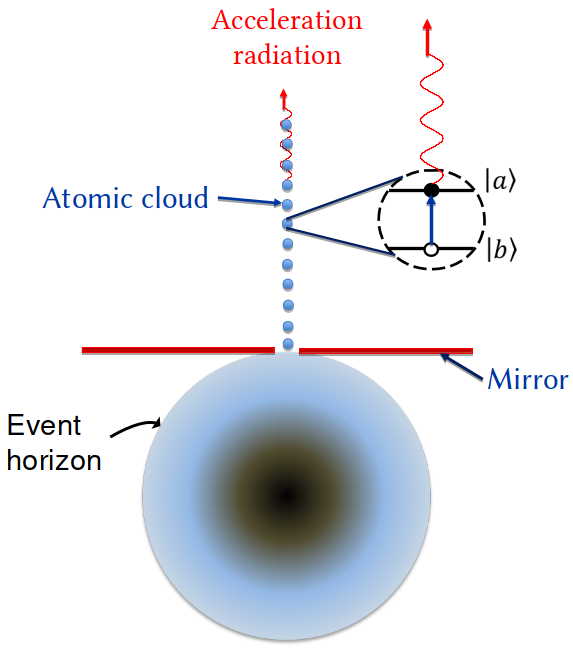}}
\caption{\label{QED}%
  Atoms in the ground state $\left\vert b\right\rangle $ are
  freely-falling in a Schwarzschild black hole metric, where the state of the
  field is the Boulware vacuum. 
  The atoms, though inertial and moving along geodesics, emit acceleration
  radiation. 
  When the atoms are released at random times from infinity, the outgoing
  field is thermal.}
\end{figure}

When Einstein first formulated the equivalence principle he was mainly
concerned with the laws of classical physics.
Ginzburg and Frolov in their review paper\cite{GinzburgFrolov} mentioned
that:  
``The question of whether or not the equivalence principle
holds for the description of phenomena for which their quantum nature is
important is by no means trivial.''

Here we discuss acceleration radiation of an atom freely-falling in the
gravitational field of a static BH.
The equivalence principle tells us that the atom essentially falls
``force-free'' into the BH, that is, the atom's acceleration is equal to
zero. 
How then could it emit something which looks like acceleration radiation?
To answer this question we consider modes of the field in the
reference frame of the black hole.
In the Schwarzschild metric the field modes are stationary, even though they
are modified by the gravitational field of the BH.
However, in the reference frame of the freely falling atom the field
modes are changing with time.

The equivalence principle is manifested as a symmetry between emission by
a static atom in Minkowski spacetime in the Rindler vacuum (discussed in
the previous section), and an atom freely falling in a gravitational field
of a BH in the Boulware vacuum. Moreover, there is an analogy between
the Rindler horizon and the BH event horizon.
Indeed, the time-radius part of Schwarzschild metric interval,
\begin{equation}
\D s^2=\frac{ r-r_{\subsc g} }r c^2 \D t^2
-\frac r{ r-r_{\subsc g} } \D r^2\,,  \label{gs1}
\end{equation}
which could be approximated near the event-horizon $r=r_{\subsc g}$ by
\begin{equation}
\D s^2\simeq\frac{ r-r_{\subsc g} }{ r_{\subsc g} } c^2 \D t^2
-\frac{ r_{\subsc g} }{ r-r_{\subsc g} } \D \bar r^2\,,
\end{equation}
and using the coordinate $\bar r$ such that
$r-r_{\subsc g}=e^{ \bar r/r_{\subsc g} }$  to describe space-time events
outside the event horizon, the time-radius Schwarzschild interval becomes
\begin{equation}
\D s^2\simeq e^{ \bar r / r_{\subsc g} }
\left( \Big.c^2 \D t^2-\D {\vec{r}\,}^2\right)\,,
\label{eq:31-46}
\end{equation}
which is the interval of the Rindler space metric, Eq.~\eqref{Ra}.
Comparing with the interval of Rindler space, Eq.~\eqref{Ra},
we obtain an effective acceleration corresponding to a free fall near the
event horizon
\begin{equation}
\Deriv {\bar r} s 2=\alpha
=\frac{1}{2r_{\subsc g}}\,.
\label{eq:09-36}
\end{equation}

Next we consider an atom launched radially from the event horizon with an
initial radial velocity $V_{0}=c\D r/\D s$ (see Fig.~\ref{QED}b).
Using the Schwarzschild metric in Eq.\ \eqref{gs1}, the equations of atomic
radial motion are  
\begin{align}
\left(\Deriv rs{} \Big.  \right)^2&=
\frac{V_0^2}{c^2}+\frac{ r_{\subsc g} }r-1\,,
\notag\\
\Deriv ts{}&=
\frac{V_{0}}{c^2\left( 1-\frac{r_{\subsc g}}{r}\right) }\,.
\end{align}%
For $V_{0}\ll c$ we find the following solution 
\begin{align}
  \frac{r}{r_{\subsc g}}
  &=1+\frac{V_{0}^{2}}{c^{2}}-\frac{s ^{2}}{4r_{\subsc g}^{2}}\,,
\notag\\
t&=\frac{r_{\subsc g}}{c}\ln \left( \frac{2r_{\subsc g}V_{0}+cs }{%
2r_{\subsc g}V_{0}-cs }\right)\,.  \label{t1}
\end{align}%
In terms of the coordinate $\bar{r}$, the atomic trajectory is 
\begin{equation}
  \bar{r}=r_{\subsc g}\ln \left[ \frac{1}{4r_{\subsc g}^{2}}
  \left( \frac{4r_{\subsc g}^{2}V_{0}^{2}}{c^{2}}-s ^{2}\right)\right]\,.
  \label{t2}
\end{equation}%
The trajectory of the atom near the BH event horizon, given by Eqs.~(\ref{t1})
and (\ref{t2}), has the same form as the trajectory of the atom fixed in
Minkowski spacetime at
\begin{equation}
z_{0}=2\frac{V_0}cr_{\subsc g}
\label{eq:11-x}
\end{equation}
viewed in the Rindler coordinates (\ref{z2a}) when relating the acceleration
in the Rindler case to the effective acceleration near the BH,
Eq.~\eqref{eq:09-36}. 
Since near the event horizon the Schwarzschild metric \eqref{gs1} can be
approximated as the Rindler metric \eqref{eq:31-46}, the probability of atomic
excitation and photon emission for an atom falling into a Schwarzschild black
hole is given by the same expressions, (\ref{t5}) and (\ref{P0}),
only where $\alpha$ and $z_{0}$ are replaced with the corresponding values,
\eqref{eq:09-36} and \eqref{eq:11-x}, respectively.

\section*{V. The ``Bogoliubov'' Path to Unruh Radiation}
In this section, we present yet another interpretation of Unruh radiation.
It could be understood as a difference of perspective between two observers.
For simplicity, we consider a real scalar field $\hat \Phi(z,t)$
to represent the photons.
This field is an operator, which can be expanded in different basis sets.
Let us consider two observers ---
a stationary and an accelerating observer (in Minkowski space)
--- which naturally have two basis sets to describe the modes of the field.
The stationary observer has the line element
$\D s^2 = c^2\D t^2 - \D z^2$, while the accelerating observer's line element
is $\D s^2 = e^{-2\bar z/\ell} ( c^2\D \bar t^2 - \D \bar z^2 )$,
which is obtained from the stationary observer's line element by transforming
to ``accelerating'' coordinates, Eq.~\eqref{z2a}.
The normal modes $\phi_\omega$ in each coordinate systems are different,
satisfying the wave equation
\begin{equation}
\frac1{ \sqrt{-g} }\partial_\mu\sqrt{-g} g^{\mu\nu}\partial_\nu\phi_\omega=0\,,
\label{eq:06-23}
\end{equation}
where $g^{\mu\nu}$ is the metric, which could be read-off from the expression
for the line element, and $g$ is its determinant.
Using Eq.~\eqref{eq:06-23}, with the metrics corresponding to Minkowski
(stationary) and Rindler \eqref{Ra} (accelerating) observers,
the normal modes for the stationary and accelerated observer
both satisfy $[(\partial_0)^2 - (\partial_1)^2]\phi = 0$,
albeit in different coordinate systems.
So in both cases the normal modes are complex exponentials,
but in terms of different coordinates.
The stationary observer's modes $\phi_\nu$, evaluated at some spacetime event, 
$(\bar z, \bar t)$ specified in Rindler coordinates, are
\begin{equation}
\phi_\nu(\bar z,\bar t)=
\frac1{ \sqrt\nu }e^{-i\nu/c \bigl( z(\bar z,\bar t)- ct(\bar z,\bar t)\bigr)}
=\frac1{ \sqrt\nu }e^{-i\nu\ell/c \, \exp[(c\bar t-\bar z)/\ell]}
\label{eq:06-24}
\end{equation}
and for the accelerating observer
\begin{equation}
\overline \phi_\nu(\bar z,\bar t)=
\frac1{ \sqrt\nu }e^{-i\nu/c (\bar z-c\bar t)}\,.
\end{equation}
So in the right Rindler wedge, the two observers describe the field as
\begin{align}
\hat \Phi(\bar z,\bar t)
&=\sum_\nu\left( \Big.\phi_\nu(\bar z,\bar t)
\hat a_\nu+\phi_\nu^*(\bar z,\bar t)\hat a_\nu^\dagger\right)
\notag\\&=\sum_{\bar\nu}\left( \Big.
\bar\phi_{\bar\nu}(\bar z,\bar t)\hat b_{\bar\nu}+
\bar\phi_{\bar\nu}^*(\bar z,\bar t)
\hat b_{\bar\nu}^\dagger\right)\,.
\end{align}
Using the orthogonality of the modes,
$\finner{ \phi_\nu(z,t) }{ \phi_{\bar\nu}(z,t) }
=\delta_{\nu,\bar\nu}$, where the inner-product is
given by Eq.~\eqref{eq:01-7}, we see that we could obtain $\hat a_\nu$'s
in terms of the $\hat b_{\bar\nu}$'s,
\begin{equation}
\hat a_\nu=
\finner{ \phi_\nu(z,t) }{\hat \Phi(z,t) }
=\sum_{\bar\nu}\left( \Big.\alpha_{\nu\bar\nu}\hat b_{\bar\nu}
+\beta_{\nu\bar\nu}\hat b_{\bar\nu}^\dagger\right)\,,
\label{eq:05-1}
\end{equation}
where $\alpha_{\nu\bar\nu}=\finner{\phi_\nu}{\bar\phi_{\bar\nu}}$,
and $\beta_{\nu\bar\nu}=\finner{\phi_\nu}{\bar\phi_{\bar\nu}^*}$.
Alternatively, one can obtain the $\hat b_{\bar\nu}$'s
in terms of the $\hat a_\nu$'s,
\begin{align}
\hat b_{\bar\nu}=\finner{\bar\phi_{\bar\nu}}{\hat\Phi}=
\sum_n\left( \Big.\alpha_{\nu\bar\nu}^* \hat a_\nu-
\beta_{\nu\bar\nu}^*  \hat a_\nu^\dagger\right)\,,
\label{eq:05-2}
\end{align}
where we have used the properties of the inner-product \eqref{eq:01-7},
\begin{align}
\finner{f}{g}=\finner{g}{f}^*=-\finner{g^*}{f^*}=-\finner{f^*}{g^*}^*\,.
\end{align}

\subsection*{Particles in the vacuum}
We can use Eq.~\eqref{eq:05-1} to make calculations,
for instance, the number of $S$ particles in the $\bar S$ vacuum is
\begin{equation}
\ave{\hat n}
=\expect{ 0_{\bar S} }{ \hat a_\nu^\dagger \hat a_\nu }{ 0_{\bar S} }
=\sum_{\bar\nu}\abs{ \beta_{\nu\bar\nu} }^2
\end{equation}
and using Eq.~\eqref{eq:05-2}, we find that the number of $\bar S$ particles
in the $S$ vacuum is
\begin{equation}
\ave{\hat {\bar n} }=
\expect{ 0_{ S} }{ \hat b_{\bar\nu}^\dagger \hat b_{\bar\nu} }{ 0_{ S} }
=\sum_{\nu}\abs{ \beta_{\nu\bar\nu} }^2\,.
\end{equation}
An interesting symmetry is that in both cases, the number of particles in the
other frame's vacuum is given by a summation of $\abs{\beta_{\nu\bar\nu}}^2$;
albeit, the two quantities involve summations over different indices.
If we use the Unruh--Minkowski modes for the modes $\phi_\nu$,
\begin{equation}
\phi_\nu(u)=
\frac{ e^{-\pi\nu\ell/2c } }{ \sqrt{\sinh(\pi\nu\ell/c)} }
\lim_{\lambda\rightarrow0^+}
\left( \Big.\frac{u-i\lambda}\ell
\right)^{i\nu\ell/c}\,,
\end{equation}
whose annihilation operator corresponds to a superposition of plane wave
$e^{i\nu'(ct-z)}/\sqrt{\nu'}$ annihilation operators $\hat \alpha_{\nu'}$,
\begin{equation}
\hat a_\nu=\Gamma\left( \Big.1+\frac{\nu\ell}c\right)
\int\frac{ \D \nu' }{ i\nu' }
\left[\Big.\frac{ ( i\nu')^{i\nu\ell/c} }{ e^{ \pi\nu\ell/c} }
	-{ (-i\nu')^{i\nu\ell/c} }\right]\hat \alpha_{\nu'}\,.
\end{equation}
$\alpha_{\nu\bar\nu}$ and $\beta_{\nu\bar\nu}$ are
\begin{equation}
\alpha_{\nu\bar\nu}=
\frac{ e^{-\frac\pi2 \nu\ell/c } }
{\sqrt{	2\sinh\left( \Big.\pi \nu\ell/c	\right)	}}
\delta_{\nu\bar\nu}\,,\quad
\beta_{\nu\bar\nu}=\frac{ e^{-\frac\pi2 \nu\ell/c } }
  {\sqrt{2\sinh\left( \Big.\pi \nu\ell/c\right)}}
\delta_{\nu\bar\nu}\,,
\end{equation}
we find that the number of Rindler photons in the Minkowski vacuum state,
and the number of Unruh--Minkowski photons in the Rindler vacuum state are both
\begin{align}
\ave{\hat n}=\ave{\hat {\bar n} }=
\frac{1}{2}\frac{1}{\exp\left( \Big.2\pi \nu\ell/c\right)-1}\,,
\end{align}
which is the Planck factor corresponding to the temperature of
$T_{\subsc U} = \hbar a/2\pi ck_{\subsc B}$.

\subsection*{An accelerating observer in Minkowski vacuum}
Notice that the Minkowski-space mode $\phi$ in Eq.~\eqref{eq:06-24}
is only defined in the right Rindler wedge, see Fig.~\ref{fig:rindwedge}. 
However, the extension to the rest of Minkowski space (into the left Rindler
wedge) is unique if we demand that it correspond to an annihilation operator, 
and that it not have any creation operator ``components'' (for all values of
the frequency parameter $\nu$). 
To correspond to an annihilation operator, it must have positive-norm,
and demanding that its norm be positive for all $\nu$, we find that it is
\begin{equation}
\phi_\nu^{\subsc R}=\frac1{ \sqrt{2\sinh(\pi\nu\ell/c)} }
\begin{cases}
e^{-\pi\nu\ell/2c}
\bar\phi_\nu^{*\subsc L}\,,&\quad \mbox{left wedge,}
\\
e^{ \pi\nu\ell/2c}\bar\phi_\nu^{\subsc R}\,,&\quad \mbox{right wedge.}
\end{cases}
\label{eq:13-26}
\end{equation}
There is another family of Minkowski modes, $\phi_\nu^{\subsc L}$,
which is concentrated mostly in the left Rindler wedge,
\begin{equation}
\phi_\nu^{\subsc L}=\frac1{ \sqrt{2\sinh(\pi\nu\ell/c)} }
\begin{cases}
e^{ \pi\nu\ell/2c}\bar\phi_\nu^{\subsc L}\,,&\quad \mbox{left wedge,}
\\
e^{-\pi\nu\ell/2c}\bar\phi_\nu^{*\subsc R}\,,&\quad \mbox{right wedge.}
\end{cases}
\end{equation}

Consider a two-level atom with constant acceleration in the right Rindler wedge,
with trajectory given by Eq.~\eqref{u1}.
In its frame, the atom interacts with the mode $\bar\phi_\nu^{\subsc R}$
in the right Rindler wedge, Eq.~\eqref{z2a},
which corresponds to the annihilation operator $\hat b_\nu^{\subsc R}$.
Thus, the time evolution of the state of the field--atom system is
given by the time-evolution operator $\hat U$ (first-order time-dependent
perturbation theory) 
\begin{equation}
\hat U\simeq
\hat 1+\frac1{i\hbar}\int_0^{\tau'}\D \tau \,
\hat \sigma^\dagger e^{i\omega\tau}\hat b_\nu^{\subsc R}e^{-i\nu\tau}\,,
\end{equation}
which means that the atomic excitation process is accompanied by
the annihilation of a a right Rindler wedge photon.

For the Minkowski observer, however, the mode which the atom interacts with
is zero in the left wedge, and he describes the annihilation operator $\hat
b_\nu^{\subsc R}$ using Eqs.~\eqref{eq:13-26} as
\begin{equation}
\hat b_\nu^{\subsc R}
=\frac1{ \sqrt{2\sinh(\pi\nu\ell/c)} }
\left( \Big.e^{ \pi\nu\ell/2c}\hat a_\nu^{\subsc R}
+e^{-\pi\nu\ell/2c}\hat a_\nu^{\dagger\subsc L}\right)\,.
\end{equation}
Thus, since $\hat a_\nu^{\subsc R}\ket{0_{\subsc M}}=0$,
the time-evolution operator, operating on the initial Minkowski vacuum state, 
is
\begin{equation}
\hat U\simeq\hat 1+
\frac1{i\hbar}\frac1{ \sqrt{2\sinh(\pi\nu\ell/c)} }
e^{-\pi\nu\ell/2c}\int_0^{\tau'}\D \tau \,\sigma^\dagger
e^{i\omega\tau}\hat a_\nu^{\dagger\subsc L}e^{-i\nu\tau}\,.
\end{equation}

\section*{VI. Periodicity Trick for Unruh Temperature}
Now we will give a ``trick'' for deriving the Unruh temperature.
The trick is to argue that, in the Rindler metric, the
time coordinate must be periodic in the imaginary direction and this
imaginary periodicity implies that Rindler spacetime has a temperature.
The original derivation of the Unruh temperature using periodicity in
imaginary time may be found in a paper by one of
us,\cite{duff,Dowker77,Dowker78}
following a similar derivation of the Hawking temperature.\cite{gibbons}

Quantum field theory at finite temperature is periodic in imaginary time,
with periodicity
\begin{equation}
t\rightarrow t+i\hbar \beta \,,  \label{f2}
\end{equation}%
where $\beta =1/k_{\subsc B}T$.
One way to see this is by looking at the thermal average, which possesses the
property  
\begin{equation}
  \left\langle \hat{Q}(t)\hat{Q}(t^{\prime })\right\rangle
  =\left\langle \hat{Q}(t^{\prime })\hat{Q}(t+i\hbar \beta )\right\rangle\,.
  \label{f3}
\end{equation}
Indeed, using the equation for the time evolution of the $\hat{Q}$ operator and
the invariance of the trace under cyclic permutation, we obtain
\begin{align}
  & \left\langle \hat{Q}(t)\hat{Q}(t^{\prime})\right\rangle
    =\frac{1}{Z}\text{Tr}\left( e^{-\beta \hat{H}}e^{i\hat{H}t/\hbar }
    \hat{Q}(0)e^{-i\hat{H}t/\hbar }\hat{Q}(t^{\prime})\right)  \notag \\
& \qquad =\frac{1}{Z}\text{Tr}\left( e^{i\frac{\hat{H}}{\hbar }\left(
t+i\hbar \beta \right) }\hat{Q}(0)e^{-i\frac{\hat{H}}{\hbar }%
\left( t+i\hbar \beta \right) }e^{-\beta \hat{H}}\hat{Q}(t^{\prime })\right)
\notag\\
& \qquad =\frac{1}{Z}\text{Tr}\left( e^{-\beta \hat{H}}\hat{Q}(t^{\prime })%
\hat{Q}(t+i\hbar \beta )\right)\,.
\end{align}
Equation~(\ref{f3}) is commonly referred to as the Kubo--Martin--Schwinger
(KMS) condition. 
Since the ordering of the field operators on the two sides are interchanged,
the corresponding periodicity along the imaginary time direction is referred
to as ``periodicity with a twist.''

Now let us assume that state of the field is the Minkowski vacuum
$\left\vert0_{\subsc M}\right\rangle$.
That is, in the inertial reference frame the temperature is equal to zero.
Then the zero temperature average over this state can be written as
\begin{equation}
  G(t,z;t^{\prime },z^{\prime })
  =\left\langle 0_{\subsc M}\right\vert \hat{\Phi}(t,z)
    \hat{\Phi}(t^{\prime },z^{\prime })\left\vert 0_{\subsc M}\right\rangle\,,
\label{f1}
\end{equation}
where $\hat{\Phi}(t,z)$ is the field operator at the spacetime event $(z,t)$.

Since the vacuum is Lorentz-invariant, the two-point function (\ref{f1})
must depend only on the Lorentz-invariant spacetime interval
$c^{2}(t-t^{\prime})^{2}-(z-z^{\prime })^{2}$.
If we make a coordinate transformation into the Rindler spacetime using
Eq.~\eqref{z2a} to express the interval in terms of the Rindler coordinates,
the average (\ref{f1}) depends on
\begin{align}
c^{2}(t-t^{\prime })^{2}-(z-z^{\prime })^{2}
&=\ell^2\left[\left(e^{\bar{z}/\ell}\sinh\left(\frac{c\bar{t}}{\ell}\right)
  -e^{\bar{z}^{\prime }/\ell}\sinh\left(\frac{c\bar{t}^{\prime }}{\ell}\right)
\right)^{2}\right. 
\notag\\
  &\hphantom{=}\mbox{}-\left.\left(e^{\bar{z}/\ell}
    \cosh\left(\frac{c\bar{t}}{\ell}\right)
  -e^{\bar{z}^{\prime }/\ell}\cosh\left(\frac{c\bar{t}^{\prime }}{\ell}\right)
	\right)^{2}\right]\,.
\end{align}
Hence, because of the periodicity of hyperbolic sine and cosine functions
under the addition of the imaginary increment $2\pi i$ to their argument, we
have
\begin{align}
  \sinh \left( \frac{c}{\ell}\bar{t}\right)&=\sinh \left[ \frac{1}{\ell}
    \left( \bar{t}+\frac{2\pi \ell}{c}i\right) \right]
  =\sinh (c\bar{t}/\ell)\cos (2\pi )\,,\notag\\
  \cosh \left( \frac{c}{\ell}\bar{t}\right)&
  =\cosh \left[ \frac{1}{\ell}\left( \bar{t}
  +\frac{2\pi \ell}{c}i\right) \right] =\cosh (c\bar{t}/\ell)\cos (2\pi)\,,
\end{align}
and we conclude that in the Rindler spacetime the two-point function
$G(\bar{t},\bar{z};\bar{t}^{\prime },\bar{z}^{\prime })$ obeys the KMS
condition, namely 
\begin{equation}
  G(\bar{t},\bar{z};\bar{t}^{\prime },\bar{z}^{\prime })
  =G\left( \bar{t}^{\prime },\bar{z}^{\prime };\bar{t}
    +\frac{2\pi \ell}{c}i,\bar{z}\right) .
\end{equation}

Comparing this with Eq.~(\ref{f3}), we see that
$2\pi\ell/c=2\pi c/\alpha=\hbar /k_{\subsc B}T_{\subsc U}$, which yields the
Unruh temperature   
\begin{equation}
T_{\subsc U}=\frac{\hbar \alpha}{2\pi ck_{\subsc B}}\,.
\end{equation}
In other words, when viewed from a uniformly accelerating frame (i.e., the
Rindler frame), the two-point function computed in the Minkowski vacuum
appears to satisfy the KMS condition (\ref{f3}).
Therefore, one may conclude that with respect to the Rindler observer,
the Minkowski vacuum looks like a thermal reservoir of temperature $T_{\subsc U}$.

\section*{VII. Conclusions}
We revisit Unruh Radiation and arrive at the effect by different means.
Using a quantum-optics route, we treat both the accelerating atom and
accelerating mirror cases, which we also treat using the Unruh--Minkowski modes.
The case of an atom freely-falling into a black hole is also discussed, and we
discuss its relation to Einstein's Equivalence Principle.
Then, we show how the effects could be obtained from Bogoliubov
transformations, and finally, we show the relation to the KMS condition, of
which Schwinger is among the namesakes.

\section*{Acknowledgments}
MOS, JSB, and AAS would like to thank the Robert A.\ Welch Foundation (Grant
No.\ A-1261), the Office of Naval Research (Award No. N00014-16-1-3054),
and the Air Force Office of Scientific Research (FA9550-18-1-0141)
for their the support.
DNP and WGU are supported by the Natural Sciences and Engineering Council of
Canada. 
MJD is supported in part by the STFC under rolling grant ST/P000762/1.
WPS thanks Texas A\&M University for a Faculty Fellowship at the Hagler 
Institute for Advanced Study at Texas A\&M University and Texas A\&M 
AgriLife  for support of this work. He is also a member of the Institute 
of Quantum Science and Technology (IQST) which is financed partially by 
the Ministry of Science, Research and Arts Baden-W\"urttemberg.

\end{document}